\documentclass[twoside]{IEEEtran}

\usepackage{cite}

\usepackage{amsmath,amssymb,amsfonts,mathrsfs,bm}

\usepackage{graphicx}
\usepackage{subfig}
\graphicspath{{./figures/}}

\usepackage{tabularx}
\usepackage{multirow}
\usepackage{booktabs}
\usepackage{threeparttable}

\makeatletter
  \newif\if@restonecol
  \def\@IEEEBIOskipN{2mm}
\makeatother

\usepackage[linesnumbered,ruled,lined]{algorithm2e}
\usepackage{algpseudocode}

\usepackage{scalerel}
\usepackage{tikz}
\usetikzlibrary{svg.path}

\definecolor{orcidlogocol}{HTML}{A6CE39}
\tikzset{
  orcidlogo/.pic={
    \fill[orcidlogocol] svg{M256,128c0,70.7-57.3,128-128,128C57.3,256,0,198.7,0,128C0,57.3,57.3,0,128,0C198.7,0,256,57.3,256,128z};
    \fill[white] svg{M86.3,186.2H70.9V79.1h15.4v48.4V186.2z}
                 svg{M108.9,79.1h41.6c39.6,0,57,28.3,57,53.6c0,27.5-21.5,53.6-56.8,53.6h-41.8V79.1z M124.3,172.4h24.5c34.9,0,42.9-26.5,42.9-39.7c0-21.5-13.7-39.7-43.7-39.7h-23.7V172.4z}
                 svg{M88.7,56.8c0,5.5-4.5,10.1-10.1,10.1c-5.6,0-10.1-4.6-10.1-10.1c0-5.6,4.5-10.1,10.1-10.1C84.2,46.7,88.7,51.3,88.7,56.8z};
  }
}

\newcommand\orcidicon[1]{%
  \href{https://orcid.org/#1}{
    \mbox{\scalerel*{
      \begin{tikzpicture}[yscale=-1, transform shape]
        \pic{orcidlogo};
      \end{tikzpicture}
    }{|}%
    }%
  }%
}

\usepackage[colorlinks,allcolors=black]{hyperref} 

\newtheorem{theorem}{Theorem}

\title{Few-Shot Testing of Autonomous Vehicles \\ with Scenario Similarity Learning}

\author{Shu Li\textsuperscript{\orcidicon{0009-0001-7579-1383}},
Honglin He\textsuperscript{\orcidicon{0000-0003-4673-5283}}, 
Jingxuan Yang\textsuperscript{\orcidicon{0000-0001-9798-7347}},
Jianming Hu\textsuperscript{\orcidicon{0000-0001-8065-7309}}, \IEEEmembership{Senior Member,~IEEE}\\ 
Yi Zhang\textsuperscript{\orcidicon{0000-0001-5526-866X}}, \IEEEmembership{Senior Member,~IEEE} and
Shuo Feng\textsuperscript{\orcidicon{0000-0002-2117-4427}}, \IEEEmembership{Member,~IEEE}

\thanks{This work is supported by Beijing Nova Program 20230484259 and Beijing Natural Science Foundation 4244092. We acknowledge the Wuhan East Lake High-Tech Development Zone (also known as the Optics Valley of China, or OVC) National Comprehensive Experimental Base for Governance of Intelligent Society. 
\textit{(Corresponding author: Shuo Feng.)}}
\thanks{Shu Li, Honlin He, Jingxuan Yang, and Jianming Hu are with the Department of Automation, Tsinghua University, Beijing 100084, China (email: li-s23@mails.tsinghua.edu.cn, hehl21@mails.tsinghua.edu.cn, yangjx20@mails.tsinghua.edu.cn, hujm@tsinghua.edu.cn).}
\thanks{Yi Zhang is with the Department of Automation, Beijing National Research Center for Information Science and Technology (BNRist), Tsinghua University, Beijing 100084, China, and also with the Tsinghua-Berkeley Shenzhen Institute (TBSI), Shenzhen 518055, China (e-mail: zhyi@mail.tsinghua.edu.cn).}%
\thanks{Shuo Feng is with the Department of Automation, Beijing National Research Center for Information Science and Technology (BNRist), Tsinghua University, Beijing 100084, China (e-mail: fshuo@tsinghua.edu.cn).}
}

\begin{document}

\maketitle

\begin{abstract}
  Testing and evaluation are critical to the development and deployment of autonomous vehicles (AVs). Given the rarity of safety-critical events such as crashes, millions of tests are typically needed to accurately assess AV safety performance. Although techniques like importance sampling can accelerate this process, it usually still requires too many numbers of tests for field testing. This severely hinders the testing and evaluation process, especially for third-party testers and governmental bodies with very limited testing budgets. The rapid development cycles of AV technology further exacerbate this challenge. To fill this research gap, this paper introduces the few-shot testing (FST) problem and proposes a methodological framework to tackle it. As the testing budget is very limited, usually smaller than 100, the FST method transforms the testing scenario generation problem from probabilistic sampling to deterministic optimization, reducing the uncertainty of testing results. To optimize the selection of testing scenarios, a cross-attention similarity mechanism is proposed to learn to extract the information of AV’s testing scenario space. This allows iterative searches for scenarios with the smallest evaluation error, ensuring precise testing within budget constraints. Experimental results in cut-in scenarios demonstrate the effectiveness of the FST method, significantly enhancing accuracy and enabling efficient, precise AV testing.
\end{abstract}


\section{Introduction}

\IEEEPARstart{T}{esting} and evaluation of autonomous vehicles(AVs) has attracted great interests of researchers in recent years\cite{li2018artificial, li2019parallel, li2022features, li2023advanced, wang2022verification}. The underlying significance of testing and evaluation for AVs arises from the safety-critical nature of open-road applications. However, the rarity of safety-critical events (e.g. crashes) within seemingly endless traffic scenarios in real world\cite{riedmaier2020survey} substantially undermines the efficacy of testing the performance index of AVs\cite{liu2022curse}. Consequently, there is a compelling imperative to judiciously generate testing scenarios to enhance the efficiency of AV evaluation processes.

Although it has been a long time since the concept of testing AVs with advanced intelligent methods is mentioned, there's still no widely accepted testing method for routine test of various kinds of AVs. We suggest that the reason lies in the fact that the available testing budget in many real-world application scenarios are extremely small, causing failure of current methods.
For example, for third-party testing organizations and governmental bodies, the generation of an extensive array of testing scenarios for all potential AV models, particularly with open road testing, is completely not realistic.
Besides, with the rapid iterative development of autonomous driving technique, conducting a thorough evaluation of AV performance within the research and development cycle also becomes increasingly infeasible. 
In these realistic cases, a preliminary yet reliable testing and evaluation result is urgently needed, and the result must be generated within the confines of an extremely small budget for testing. The testing procedure should also be quite concise and deterministic, thus being able to expediently generalize among numerous possible AVs under test.
Moreover, the quantitative and explainable results are needed as a benchmark for comparing the performance of diverse AVs, which creates additional difficulties. 

In this paper, we formulate this problem as the general ``few-shot testing" (FST) problem and propose the FST method to tackle this problem. To the best of our knowledge, this is the first time the general FST problem is developed and general FST method is discussed. 


\begin{figure}[!t]
  \centering
  \includegraphics[width=8.85cm]{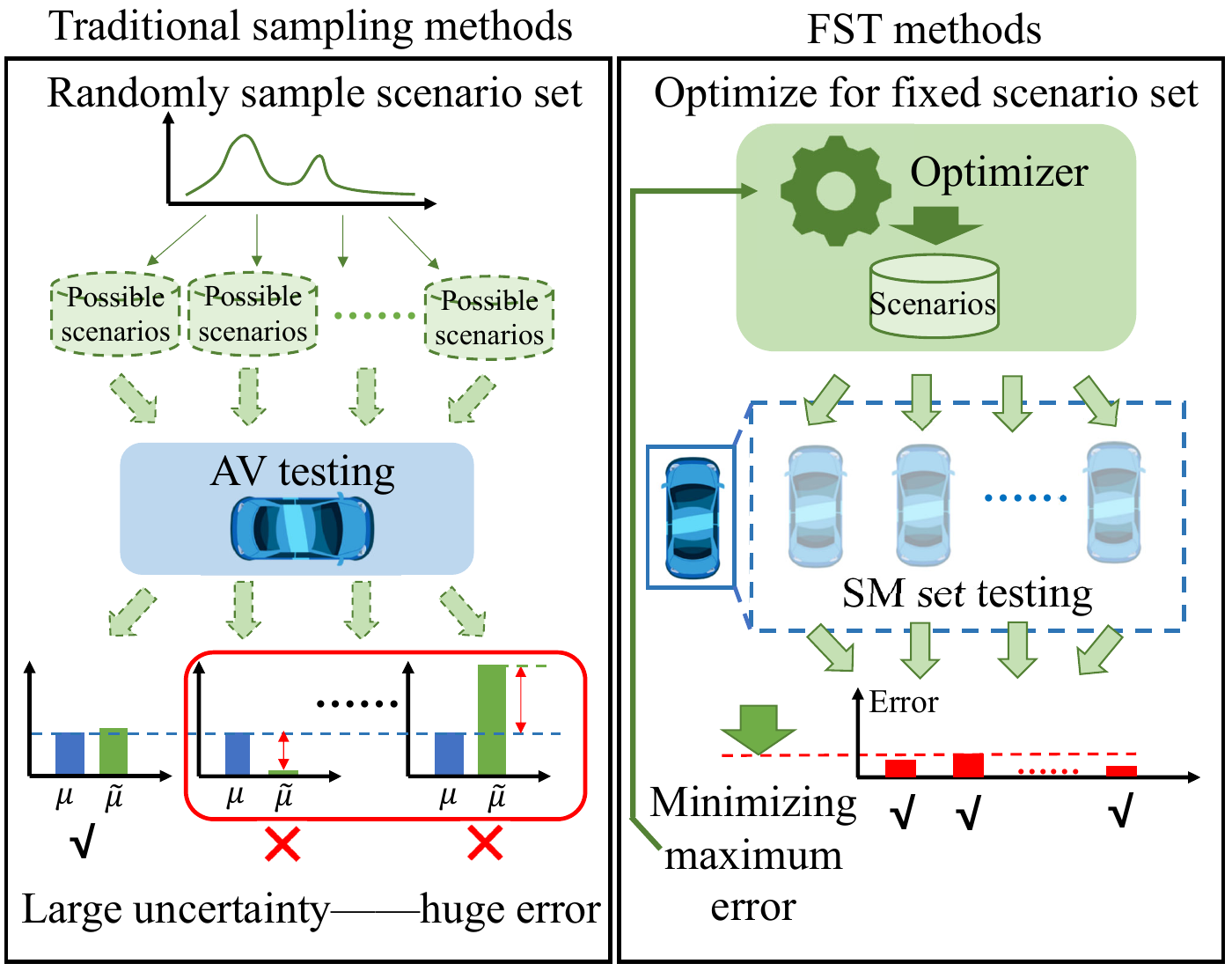}
  \caption{The basic idea of FST method is using determinate scenarios to generalize among AVs and reach precise evaluation results.}
  \label{fig:idea}
\end{figure}

 Towards solving FST problem, current testing methods failed because of unable to quantify the performance index of AVs or unable to control the substantial uncertainty and testing variance with such a small testing budget (e.g. small than $10^2$ scenarios for test).
 In the light of these failures, we apply statistical models to quantify the performance index of AVs. Furthermore, to eliminate the uncertainty caused by statistical sampling method, we transform the testing problem from a probabilistic sampling problem to a deterministic optimization problem, as shown in Fig.~\ref{fig:idea}. In the deterministic optimization process, we search for a FST scenario set with best generalization ability to improve the applicability of FST method in real world test to different AVs. To leverage the limited available testing budget, the FST scenario set is searched from a global perspective but not sampled sequentially. Finally, with the fixed scenario set, we use fixed evaluation error to evaluate the precision of FST method. With some information about AVs, a minimum upper bound of evaluation error can be generated to ensure the reliability of FST scenarios.

Specifically, we apply surrogate models (SMs, as in \cite{yang2023adaptive, feng2020part2}) as a representation of possible AVs under test. To extract the information from the scenario space of AVs, we propose a cross-attention similarity network to learn the relationships between selected FST scenarios and other scenarios. The similarity network serves as a measurement of global information gain of testing scenarios and is used for the fusion of AV's testing results to get final evaluation result. After finishing training the similarity network, we use gradient descent optimizer to globally and iteratively search for a FST scenario set with best generalization ability and minimized upper bound of evaluation errors. Eventually, we are able to generate optimized scenario utilization strategy with optimized FST scenarios.

We designed the experiments on cut-in case (commonly used in previous studies \cite{zhao2017accelerated, zhang2018accelerated, feng2020part2} on AV testing) to test and evaluate the performance index of AVs with extremely small number of scenarios ($n=5, 10, 20$). To the best of our knowledge, this small number of testing scenarios has never been applied in other works. The results of our experiment show that proposed FST method markedly outperforms all current methods on accuracy metrics. The relative error of FST method can still be controlled in an acceptable small scope even if testing budget is strictly limited, rendering an unprecedented possibility of quick and accurate test for AVs.

The contribution of this paper can be summarized as follows:

(1) We formulate the general FST problem and comprehensively analyze the features and underlying challenges of FST problem compared to traditional testing and evaluation problems.

(2) We propose the FST method and theoretically analyze its optimality, accuracy and sensitivity. The performance index of AV can be evaluated with certain upper bound of estimation error. With a novel and generalized deep learning framework, FST method is capable of learning to extract the scenario space similarity according to the features of AV under test.

(3) We provide abundant experiments to verify the effectiveness of FST method from different perspectives and the experiments show a significant improvement of testing and evaluation accuracy compared with previous methods. 

\section{Related Works}
Many efforts have been made to search for a smaller testing scenario set or accelerate testing process from different perspectives. Although they cannot be applied directly in FST cases, we would still give a review of these methods and analyze their limitations.

Using critical or risky scenarios to test AVs is intuitively easier to discover the defects of AV and reduce cost of tests.
As a practical method, some autonomous driving companies maintain a scenario set from logged data and expert knowledge to verify the reliability of their AVs before on-road deployment\cite{webb2020waymo}. Searching for critical scenarios or corner cases is also a commonly used scheme to generate a smaller testing scenario set\cite{zhang2022finding}. Based on knowledge\cite{bagschik2018ontology}, scenario clustering\cite{kruber2018unsupervised, kruber2019unsupervised}, scenario coverage\cite{weissensteiner2023operational}, optimization strategy\cite{duan2020test, klischat2019generating} or other carefully designed models\cite{li2021scegene, li2016intelligence, ge2024task, zhang2018accelerated}, many methods are capable of generating a representative scenario set with certain risks. Deep learning\cite{pronovost2023scenario} and reinforcement learning\cite{sun2021corner} methods are also effective tools for scenario generation. Given a small testing budgets, these methods will be able to generate certain testing scenarios. However, the efficiency of these scenarios is usually measured by the risk, realism, or other specifically designed metrics. Apart from being unable to quantify the performance index of AVs, it is still doubtful whether these metrics are persuasive enough for a reliable evaluation result of AVs.

Statistical sampling methods represent an effective approach to quantifying the performance index of AV model while generating critical scenarios to accelerate the testing process\cite{feng2020part1, feng2020part2, feng2021intelligent, feng2023dense, zhao2016accelerated, zhao2017accelerated}. Based on naturalistic driving data (NDD), the naturalistic driving environment (NDE) can be constructed and is widely used in these statistical testing methods. Furthermore, the performance of AVs can be estimated with a critical distribution using importance sampling (IS). These methods can generate unbiased quantitative results with higher efficiency. However, the critical scenarios may be similar or repetitive and is possible a waste of AV information. Additionally, controlling the testing variance in a extremely smaller test set becomes difficult due to uncertainty and thereby the effectiveness of these method is greatly impaired in FST cases.

The above methods generate critical or risky scenarios according to off-line strategies. By leveraging bayesian optimization or regression methods, it is also possible to adaptively testing the performance index of AV, thus utilize potential information of AVs \cite{gangopadhyay2019identification, feng2020testing, yang2023adaptive, yang2024adaptive}. However, in FST cases where the testing budget is extremely small, the extra information of AV is not enough for a significantly improved testing and evaluation result. Meanwhile, as these method have to generate scenarios incrementally, some global information of AV may be missed and the FST accuracy is hard to be ensured.

In \cite{li2024few}, we proposed a coverage-based FST method (FST-C). With the handcrafted coverage model, the testing precision can be greatly improved compared with previous methods. However, the effectiveness is highly relevant to the design of coverage model and AV under test, which may be inapplicable to various AVs or different application scenarios. In many specific cases, the error of the handcrafted method is not small enough for a useful and accurate testing and evaluation result. The target of general few-shot testing method applicable to diverse AVs and various scenarios is still not reached.

\section{Problem Formulation}

\subsection{Performance Index Testing}

In order to quantify the performance index (e.g. crash rate) of AV under test, we use the NDE to model the driving environment of AVs, which is a general formulation and is applied in many existing studies{\cite{feng2020part1, feng2021intelligent, feng2023dense}}. In NDE, the testing state space $\mathcal{X}$ is restricted by the operational design domain. Then the exposure frequency and the testing performance of AV on certain scenario $\bm{x}\in\mathcal{X}$ is defined as $p(\bm{x})$ and $P(A|\bm{x})$. We consider $P(A|\bm{x})$ as the probability of event of interest $A$ (e.g. crashes) on scenario $\bm{x}$. Then $P(A)$, which means the overall performance index of AV and also the overall probability of event $A$, is defined as follows:
\begin{equation}\label{eq:nde_gt}
    P(A)=\sum_{\bm{x}\in\mathcal{X}}{P(A|\bm{x})p(\bm{x})}.
\end{equation}

Viewing from the perspective of random variable $\bm{X}$ taking concrete values $\bm{x}\in\mathcal{X}$ with probability distribution $p(\bm{x})$, Eq.~(\ref{eq:nde_gt}) can also be written as
\begin{equation}
    \mu = \mathbb{E}_p[P(A|\bm{X})],
\end{equation}
where $\mu=P(A)$ is the mathematical expectation of $P(A|\bm{X})$ and serves as the ground truth of the performance index of AV under test.

For traditional methods that test AV directly in NDE, testing scenarios are generated with Crude Monte Carlo (CMC)\cite{mcbook}. CMC method samples a testing scenario set $\mathcal{X}_n \triangleq \bm{x}_1, ..., \bm{x}_n$ from the original distribution $p(\bm{x})$, and the expectation of $P(A|\bm{X})$ is estimated by

\begin{equation}\label{eq:target_cmc}
    \Tilde{\mu}_{\mathrm{CMC}} = \frac{1}{n}\sum_{i=1}^{n}{P(A|\bm{x}_i)},\bm{X}_i\sim p(\bm{X}).
\end{equation}
As $n\rightarrow \infty$, it can be proved that $\Tilde{\mu}_{\mathrm{CMC}}$ converges to $\mu$ with the probability of 1. Therefore CMC is unbiased and can be used as a reliable estimation of AV performance index with a sufficiently large number of tests. According to reports \cite{RR-2662}, millions or even billions of miles of test are needed to demonstrate the reliability of AVs in fatal crashes and this is almost impossible in real world application. As a contrast, when the number of scenarios $n$ is limited within a very small range (such as $n \leq 10^2$), the estimation variance of CMC is almost impossible to be controlled.

\subsection{Target of Few-Shot Testing}

The fundamental testing target of FST problem centers on testing and evaluating the performance index of AV with a strictly restricted number of tests $n$. In this study, we tackle the testing problem with the idea of optimizing instead of sampling from distributions. Because of the substantial uncertainty when $n$ is a small number, sampling methods usually result in a large variance. By contrast, we search for a fixed testing scenario set $\mathcal{X}_s$ given specific $n$ and try to achieve smallest evaluation error comparing with the ground truth of AV performance:

\begin{equation}\label{eq:target_fst}
    \mathop{\min}_{\mathcal{X}_n} E=\left|\Tilde{\mu}_{\mathrm{FST}}-\mu\right|.
\end{equation}

In Eq.~(\ref{eq:target_fst}), $\Tilde{\mu}_{\mathrm{FST}}$ is a fixed value after the testing set $\mathcal{X}_n$ is determined. Then the evaluation error will be a certain value with no variance. With this scheme, we transform the problem of minimizing the variance with unbiased sampling method into minimizing the certain errors with fixed testing scenarios. Compared with statistical sampling method, the advantages of this conversion when $n$ is extremely small mainly lie in:

(1) with fixed and optimized FST set, the uncertainty is eliminated, which ensures the accuracy and reliability of FST method in cases where the testing budget is strictly limited;

(2) all scenarios are selected jointly with a high-level strategy but not generated sequentially or independently from distributions, and this maximizes the utility of each testing scenario from a global perspective.

\subsection{General Few-Shot Testing Problem}
Generally, the estimation result of FST method is a function of testing scenarios and we can rewrite Eq.~(\ref{eq:target_fst}) with

\begin{equation}\label{eq:mu_fst}
    \Tilde{\mu}_{\mathrm{FST}} = f\left[P(A|\bm{x}_1),\dots,P(A|\bm{x}_n)\right]
\end{equation}
to get the expanding form of FST target

\begin{equation}\label{eq:target_fst_2}
    \mathop{\min}_{\mathcal{X}_n} E=\left|f\left[P(A|\bm{x}_1),\dots,P(A|\bm{x}_n)\right]-\mu\right|.
\end{equation}

In Eq.~(\ref{eq:target_fst_2}), however, it is impossible to get all information on AV (namely $P(A|\bm{x}_i)$ and $\mu$) before testing to solve this optimization problem.
In practical terms, only part of the prior-knowledge on AV model under test is known. Moreover, since the FST set is fixed after optimization and the AV under test is unknown, FST method must have a strong generalizaiton ability across potential AVs. Consequently, we suppose that the possible AV models forms a vehicle model set $\mathcal{M}$. For all possible models $m\in\mathcal{M}$, the performance index of AV in scenario $\bm{x}$ can be tested as $P_m(A|\bm{x})$. The set $\mathcal{M}$ can be considered as an AV model with some noise and uncertainty or a number of possible AV models for test. As FST method is designed to give quick tests with minimized errors, we can further formulate Eq.~(\ref{eq:target_fst_2}) as

\begin{equation}\label{eq:target_fst_3}
    \mathop{\min}_{\mathcal{X}_n}\mathop{\max}_{m\in\mathcal{M}} E=\left|f\left[P_m(A|\bm{x}_1),\dots,P_m(A|\bm{x}_n)\right]-\mu_m\right|,
\end{equation}
where $\mu_m = \mathbb{E}_p[P_m(A|X)]$ is the ground truth of performance index for specific model $m$ and $E$ is testing and evaluation error. 

We can see from Eq.~(\ref{eq:target_fst_3}) that $n$ scenarios are carefully selected to extract the performance of all possible AVs under test. Supposing the real AV under test $m^*$ satisfies $m^*\in\mathcal{M}$, the accuracy and reliability of FST results are ensured by an upper bound of error, thus resolving the problem of substantial uncertainty and unacceptable confidence levels brought by extremely small $n$. With this general form of FST problem, the AV is hopefully to be tested with strictly limited budgets to get high accuracy.

Accordingly, the challenges to solving this problem can be summarized as follows:

(1) the form of estimation function $f$ is highly flexible and the method to generate $\Tilde{\mu}_{\mathrm{FST}}$ with testing performances $P(A|\bm{x}_i), i=1,...,n$ remains undetermined;

(2) supposing a concrete $f$ is decided, a set of scenarios $\mathcal{X}_n$ should be carefully selected to obtain a minimized upper bound of evaluation error;

(3) if $\mathcal{M}$ contains additional error and we don't have $m^*\in\mathcal{M}$, the effectiveness of testing and evaluation result might be impaired.

Our solutions to these challenges will be discussed in the next section.

\section{Few-Shot Testing Method}\label{sec:method}

\subsection{Vehicle Model Set Construction}
As a representation of the prior-knowledge on possible AVs, we use the surrogate model (SM) to construct $\mathcal{M}$ in Eq.~(\ref{eq:target_fst_3}) in this paper. As in \cite{feng2020part2, yang2023adaptive}, SM is usually an effective way to draw a sketch on AV models for further testing and evaluation. For FST problem, we focus on the generalization ability on AV with some uncertainty or multiple AVs to get minimized error, so we use multiple SMs $m_1,...,m_s$ to form a surrogate model set as $\mathcal{M}$. For simplicity, we supposed that

\begin{equation}\label{eq:sm}
    \mathcal{M} \triangleq \left\{m|m=\sum_{i=1}^{s}{c_i m_i}, c_i>0, \sum_{i=1}^{s}{c_i}=1\right\},
\end{equation}
which means the AV model under test $m^*$ can be approximated with a linear combination of $s$ possible SMs (the combination is in the performance level, i.e. $P_m(A|\bm{x})$). These SMs could consist of vehicle models from aggressive driving policies to conservative driving policies and can depict the possible strategies of AVs. Eq.~(\ref{eq:sm}) is not the only possible description of $M$. For instance, SM with noises of different magnitudes in the scenario space can also form a set $M$ which reflect the confidence we have on the prior-knowledge of AVs and this could be discussed in the future. 

With this simple form of SM set, the optimization problem in Eq.~(\ref{eq:target_fst_3}) will be easier to solve and we have the following theorem:
\begin{theorem}\label{thm:linear_simplification}
  The following 2 descriptions of minimax problem are equivalent under Eq.~(\ref{eq:sm}):
  \begin{subequations}
  \begin{align}
    \mathop{\min}_{\mathcal{X}_n}\mathop{\max}_{m\in\mathcal{M}} &\left|f\left[P_m(A|\bm{x}_1),\dots,P_m(A|\bm{x}_n)\right]-\mu_m\right| \label{eq:equivalent_optimality_form_a}, \\
    \mathop{\min}_{\mathcal{X}_n}\mathop{\max}_{i=1,...,s} &\left|f\left[P_{m_i}(A|\bm{x}_1),\dots,P_{m_i}(A|\bm{x}_n)\right]-\mu_{m_i}\right|.\label{eq:equivalent_optimality_form_b}
  \end{align}
  \end{subequations}
\end{theorem}
    
\begin{IEEEproof}
  For fixed $\mathcal{X}_n$, we focus on the maximization problem. Assuming that the optimal solution of the maximization problem in Eq.~(\ref{eq:equivalent_optimality_form_a}) is $\bm{c}^*=(c_1^*, ..., c_s^*)$ with optimal value $E^*$ and there exists $j\neq k$ such that $0<c_j^*, c_k^*<1$. For another two solution $\bm{c}'$ and $\bm{c}''$, taking 
  \begin{equation*}
  \begin{aligned}
      c'_i = c''_i &= c_i^*, \forall i \neq j,k, \\
      c_j'=c_k''&=c_j^*+c_k^*, \\
      c_k'=c_j''&=0,
  \end{aligned} 
  \end{equation*}
  and denoting $E'$, $E''$ as the objective function value respectively, we have
  \begin{equation*}
  \begin{aligned}
      E^* = \frac{c_j^*}{c_j^*+c_k^*}E'+\frac{c_k^*}{c_j^*+c_k^*}E''\leq \max\{E', E''\}.
  \end{aligned} 
  \end{equation*}
  Thus $\bm{c}^*$ is not the optimal solution and $c_j^*$ or $c_k^*$ can be converted to 0 to reach $\bm{c}'$ or $\bm{c}''$. Repeat this operation and the optimal value is possible to be taken only at $\bm{c}^*$ where $c_i^*=1, c_j^*=0, \forall j\neq i$, which is the form in Eq.~(\ref{eq:equivalent_optimality_form_b}).
  
\end{IEEEproof}
Consequently, with this construction of SM set, the complex optimization problem in Eq.~(\ref{eq:target_fst_3}) is simplified. This will be useful when we solve this optimization problem in experiments.

\subsection{Testing Results Fusion}

In this section, we try to solve the problem of fusing the testing results on AV, namely $P(A|\bm{x}_i), ..., P(A|\bm{x}_n)$, to evaluate $\Tilde{\mu}_\mathrm{FST}$ (see Eq.~(\ref{eq:mu_fst})). Classic testing method CMC provide us with a direct way in Eq.~(\ref{eq:target_cmc}), where all scenarios sampled from distribution $p(\bm{x})$ are assigned equal weight $1/n$. In this paper, we extend this form to a more general weighted sum to evaluate the contribution of each FST scenario (the testing results) to the evaluation result
\begin{equation}\label{eq:weighted_sum1}
\begin{aligned}
    \Tilde{\mu}_{\mathrm{FST}} &= f[P(A|\bm{x}_i), ..., P(A|\bm{x}_n)]\\
    &= \sum_{i=1}^{n}{P(A|\bm{x}_i)w(\bm{x}_i; \mathcal{X}_n)},
\end{aligned}    
\end{equation}
where $w(\bm{x}_i; \mathcal{X}_n)$ is the assigned weight to each testing result and is related to all testing scenarios $\mathcal{X}_n$. In order to get a normalized evaluation result $\Tilde{\mu}_{\mathrm{FST}}$, we suppose that
\begin{equation}\label{eq:weighted_sum2}
    \sum_{i=1}^{n}{w(\bm{x}_i; \mathcal{X}_n)} = 1.
\end{equation}
 Comparing with statistical methods like CMC and IS\cite{zhao2016accelerated, zhao2017accelerated, feng2020part1}, which suffer from uncertainty and extremely large variance given a small number $n$, our testing results fusion strategy is more adaptable to FST cases. As $w$ is a flexible value in $[0,1]$, the contribution of each scenario to the evaluation result is adjusted by the information it contains, thus ameliorating the problem of information redundancy or insufficiency for statistical testing methods. Additionally, $w$ is a function of the whole testing set $\mathcal{X}_n$. The weight of each scenario is determined jointly instead of independently, which maximizes the utilization of the small number of testing scenarios.

In the FST problem formulation, we aimed at finding the optimized scenario set with fixed and smallest error. Consequently, the optimality of testing result fusion strategy is important for FST method. For FST method defined with Eq.~(\ref{eq:weighted_sum1}-\ref{eq:weighted_sum2}), we can prove it in the following theorem that for any possible AV model $m$, there exists a weight function $w$ and FST set $\mathcal{X}_n$ so that the evaluation error on $m$ is 0.

\begin{theorem}\label{thm:optimality}
  Let $\mu$ be the performance index of model $m$ and $\mathcal{X}_n$ be the FST scenario set of $n$. $\Tilde{\mu}_{\mathrm{FST}}$ is given by Eq.~(\ref{eq:weighted_sum1}). Then there exists a weight function $w(\bm{x}_i; \mathcal{X}_n)$ satisfying Eq.~(\ref{eq:weighted_sum2}) so that
  \begin{equation}\label{eq:optimality}
    \Tilde{\mu}_{\mathrm{FST}} = \mu.
  \end{equation}
\end{theorem}
    
\begin{IEEEproof}
  Given fixed $\bm{x}_i$ and $P(A|\bm{x}_i)$, we can write $\Tilde{\mu}_{\mathrm{FST}}$ as
  \begin{equation*}
  \begin{aligned}
      \Tilde{\mu}_{\mathrm{FST}} &= \sum_{i=1}^{n}{P(A|\bm{x}_i)w(\bm{x}_i; \mathcal{X}_n)},\\
      &= g(w_1, ...,w_n),
  \end{aligned}
  \end{equation*}
  where $w_i = w(\bm{x}_i; \mathcal{X}_n)$ for short. According to the definition of $\mu$ we have $\mathop{\min}_{i=1,...,n}P(A|\bm{x}_i) \leq \mu \leq \mathop{\max}_{i=1,...,n}P(A|\bm{x}_i)$. Then define $j=\mathop{\mathrm{argmin}}_{j}{P(A|\bm{x}_j)}$ and $k=\mathop{\mathrm{argmax}}_{k}{P(A|\bm{x}_k)}$. Given $w_j=1$ and $w_i=0, \forall i\neq j$, we have 
  \begin{equation*}
      \Tilde{\mu}_{\mathrm{FST}} = g(w_1, ...,w_n) = P(A|\bm{x}_j) \leq \mu
  \end{equation*}
  and given $w_k=1$ and $w_i=0, \forall i\neq k$, we have
  \begin{equation*}
      \Tilde{\mu}_{\mathrm{FST}} = g(w_1, ...,w_n) = P(A|\bm{x}_k) \geq \mu. 
  \end{equation*}
  
  With the arbitrariness of $w_1, ...,w_n$ we have the continuity of $g(w_1, ...,w_n)$ and Eq.~(\ref{eq:optimality}) can be acquired.
\end{IEEEproof}

With Theorem~\ref{thm:optimality}, it is significant to find a weight function to achieve Eq.~(\ref{eq:optimality}). Following the insight of extracting as much information as possible through FST scenario set, we propose the similarity measurement as an effective representation of the information of scenarios, denoted as
\begin{equation}\label{eq:weight_function}
    w(\bm{x}_i; \mathcal{X}_n) = \sum_{\bm{x}\in\mathcal{X}}{S(\bm{x}_i, \bm{x}; \mathcal{X}_n)p(\bm{x})},
\end{equation}
where $S(\bm{x}, \bm{x}'; \mathcal{X}_n)$ is a function of similarity measurement between scenario $\bm{x}$ and $\bm{x}'$ in scenario space $\mathcal{X}$. As we calculate $w$ from a global perspective according to not only $\bm{x}_i$ but also $\mathcal{X}_n$, the similarity measurement is also relevant to both central scenarios $\bm{x}, \bm{x}'$ and FST scenario set $\mathcal{X}_n$. This enables a better evaluation accuracy with extremely small number of scenarios.

In Eq.~(\ref{eq:weight_function}) the weight of the FST scenario $\bm{x}_i\in \mathcal{X}_n$ is interpreted as the summation of its similarity to each scenario in the state space, weighted by the exposure frequency. If the similarity measurement satisfies 
\begin{equation}{\label{eq:similarity_normalization}}
    \sum_{i=1}^{n}{S(\bm{x}_i, \bm{x}; \mathcal{X}_n)} = 1, \forall \bm{x}\in\mathcal{X}, 
\end{equation}
combining with the fact $\sum_{\bm{x}\in\mathcal{X}}{p(\bm{x})}=1$ and substituting Eq.~(\ref{eq:similarity_normalization}) into Eq.~(\ref{eq:weight_function}), we can know that this similarity measurement satisfies our normalization condition in Eq.~(\ref{eq:weighted_sum2}).

Intuitively, it reasonable to decide the contribution of each FST scenario according to the number of other possible scenarios that are similar to it. The concept of similarity is also used by many methods of AV testing\cite{lou2022testing} or scenario clustering\cite{riedmaier2020survey}. Usually, similarity is a pre-defined handcrafted model (e.g. distance in the state space) and is fixed in the testing process. As similarity is a reflection or hypothesis on the features of the state space, its effectiveness is strongly relevant to the specific AV under test. Therefore, the reliability of handcrafted model is in question.

\subsection{Scenarios Similarity Learning}\label{sec:fst_similarity_learning}

\begin{figure*}[htb]
  \centering
  \includegraphics[width=18cm]{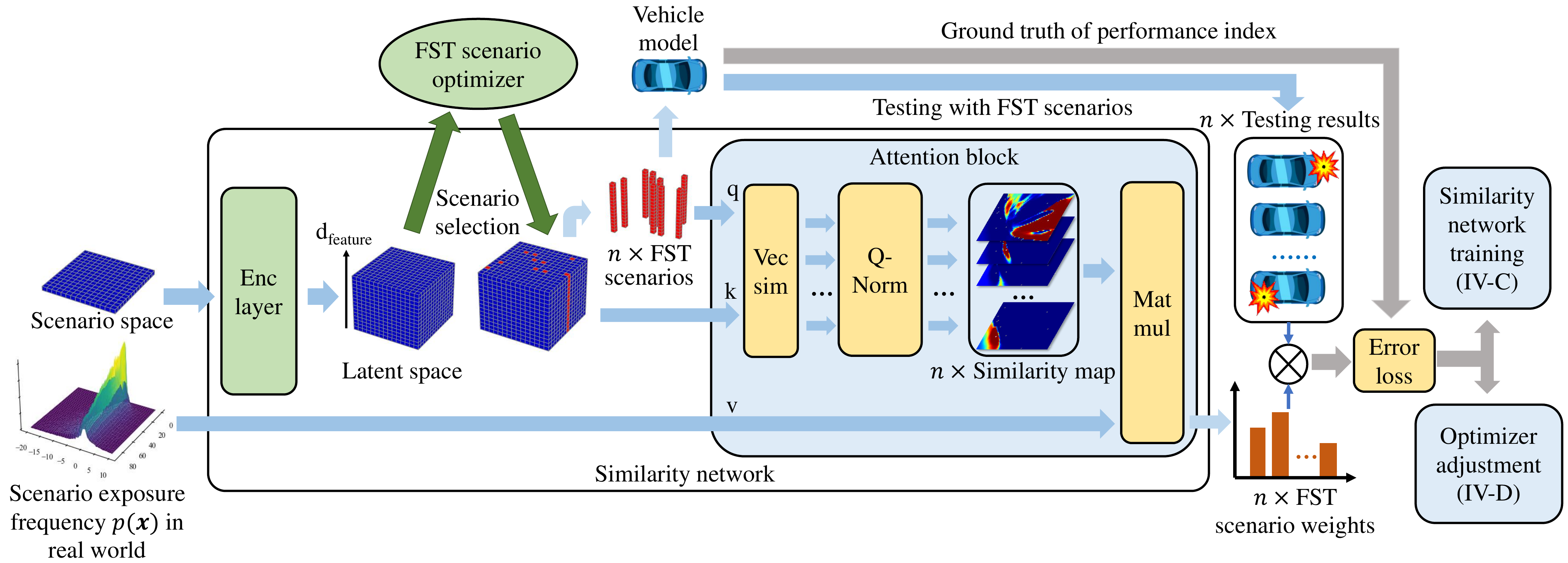}
  \caption{The structure of cross-attention similarity network and FST framework.}
  \label{fig:fst_similarity_framework}
\end{figure*}

Based on the testing result fusion strategies from Eq.~(\ref{eq:weighted_sum1}) to (\ref{eq:similarity_normalization}), we propose a general learnable similarity measurement to extract the structure in state space for an accurate testing and evaluation result.
Our purpose is to design a similarity function $S(\bm{x}_i, \bm{x};\mathcal{X}_n)$ satisfying Eq.~(\ref{eq:similarity_normalization}) given the FST scenario set and common scenario $\bm{x}$ so that the upper bound of testing and evaluation error $E$ in Eq.~(\ref{eq:target_fst_3}) is minimized. For the purpose of leveraging the possible data brought by SMs, we propose a cross-attention similarity network in a self-supervised learning manner. The learning framework is shown in Fig.~\ref{fig:fst_similarity_framework}.

Instead of applying similarity model directly on different states, we use the MLP to extract the latent features of all possible scenarios according to the vehicle model set $\mathcal{M}$. The key to the similarity network is generating a similarity measurement between a scenario in the FST scenario set $\bm{x_i}\in\mathcal{X}_n$ and a scenario in scenario space $\bm{x}\in\mathcal{X}$. This similarity value is supposed to be adjusted according to $\mathcal{X}_n$ to make use of global information. Therefore, regarding each scenario as a token, we propose the cross attention structure between FST scenarios $\bm{x}_i$ and common scenarios $\bm{x}$. All encoded samples in FST scenarios set $\bm{x}_i$ are used as queries to compute attention to other encoded scenarios $\bm{x}$, which serve as keys. In order to dynamically adjust the attention mechanism according to all of our queries $\mathcal{X}_n$, we apply the query-level normalization, which normalizes the weights of values assigned to queries along the dimension of queries. This normalization strategy extract the relative information of FST scenario set $\mathcal{X}_n$. The similarity matrix is generated as
\begin{equation}\label{eq:similarity_matrix}
    \bm{S} = \mathrm{Norm}_q\left\{d(\bm{Q},\bm{K})\right\},
\end{equation}
where $\bm{Q}_{n\times d}$ and $\bm{K}_{N\times d}$ is the encoded feature matrix from $\mathcal{X}_n$ and $\mathcal{X}$. $N$ is the number of scenarios in state space. $d(\bm{Q},\bm{K})$ is a feature-wise similarity calculation strategy (e.g. $\bm{Q}\bm{K}^T$ for classic attention). Here we compute the reciprocal of L2 distance between each column in $\bm{Q}$ and $\bm{K}$ to get $d(\bm{Q},\bm{K})_{n\times N}$. Softmax is adopted as the normalization function along the dimension of queries and the element $s_{ij}$ in $\bm{S}_{n*N}$ means the similarity between the $ith$ FST sample and the $jth$ scenario in scenario space, i.e. $S(\bm{x}_i, \bm{x}_{(j)};\mathcal{X}_n)$. Finally, according to Eq.~(\ref{eq:weight_function}), we merge $p(\bm{x})$ to get matrix $\bm{V}_{N\times 1}$ and use the attention out put of FST scenario tokens as the weight function
\begin{equation}\label{eq:weight_matrix}
    \bm{W}_{n\times 1}=\bm{S}_{n\times N}\bm{V}_{N\times 1},
\end{equation}
and $\bm{W} = \left[w(\bm{x}_i;\mathcal{X}_n), ...,w(\bm{x}_n;\mathcal{X}_n)\right]^T$. With the cross attention mechanism, the weight $w$ and estimation result $\Tilde{\mu}_{FST}$ can be calculated for the similarity network training. Note that with continuous similarity network parameters $\bm\theta$ and the query-level normalization, Eq.~(\ref{eq:similarity_normalization}) is satisfied so the optimal testing and evaluation error is 0 for specific vehicle model $m$ with optimal ${\theta}$, according to Theorem~\ref{thm:optimality}. 

After obtaining weight matrix in Eq.~(\ref{eq:weight_matrix}) we can calculate FST result according to result fusion strategy in Eq.~(\ref{eq:weighted_sum1}) and calculate further the upper bound of evaluation error in Eq.~(\ref{eq:target_fst_3}). We directly apply the upper bound of error as the loss function. As the testing result fusing function $f$ is derived from our similarity network, we regard it as $f_\theta$ and have
\begin{equation}\label{eq:loss_function}
    L(\bm{\theta}, \mathcal{X}_n) = \mathop{\max}_{m\in\mathcal{M}}\left|f_{\bm{\theta}}\left[P_m(A|\bm{x}_1),\dots,P_m(A|\bm{x}_n)\right]-\mu_m\right|.
\end{equation}
Generally the similarity network is trained by randomly sampled $\mathcal{X}_n$ to optimize $\theta$ and get minimized expectation of loss $\mathbb{E}_{\mathcal{X}_n}[L(\bm{\theta}, \mathcal{X}_n)]$. This would minimize the testing and evaluation error on all FST sets $\mathcal{X}_n$. However, if we consider the optimization of $\mathcal{X}_n$ in Eq.~(\ref{eq:target_fst_3}), our optimization target will be

\begin{equation}\label{eq:loss_training_ideal}
    \mathop{\min}_{\bm{\theta},\mathcal{X}_n} L(\bm{\theta}, \mathcal{X}_n),
\end{equation}
which means we should optimize $\bm{\theta}$ and $\mathcal{X}_n$ simultaneously.

Eq.~(\ref{eq:loss_training_ideal}) can be interpreted as prioritizing the loss value at the optimal point $\mathcal{X}_n^*$ rather than closely monitoring the average loss value on other FST scenarios.

However, $\mathcal{X}_n^*$ is completely unknown before training and the optimization of $\mathcal{X}_n$ is coupled with the optimization of $\bm\theta$, rendering Eq.~(\ref{eq:loss_training_ideal}) an intractable problem. We simplify and decouple this problem into a two-stage process for similarity network $\bm{\theta}$ training and FST set $\mathcal{X}_n$ optimizing. In the training process, as $\mathcal{X}_n^*$ is unknown, we use a compromised and practical form to approximate it by sampling $\mathcal{X}_n$ from some critical distribution, written as
\begin{equation}\label{eq:loss_training_practical}
    \mathop{\min}_{\bm{\theta}}\mathbb{E}_{\mathcal{X}_n\sim P_c}[L(\bm{\theta}, \mathcal{X}_n)],
\end{equation}
where $P_c$ is the critical distribution. It is supposed to be close to the optimal solution $\mathcal{X}_n^*$. With this strategy, the effectiveness of similarity network can be estimated directly with the forward propagating for testing and evaluation error. The training is also efficient with self-supervised manner. In this paper, we apply k-means clustering on the state space according to the performance of SMs to get several subsets of $\mathcal{X}$, denoted as $\mathcal{X}_{(i)}$. Scenarios in $\mathcal{X}_n$ are uniformed sampled from $\mathcal{X}_{(i)}$. It means that the FST scenario set used for training contains the information of SMs in different areas of the state space. Therefore the training scenario set is possible to be an effective approximation to optimal $\mathcal{X}_n^*$ and is used as $P_c$ to improve the training performance.

After the training stage, $L(\bm{\theta}, \mathcal{X}_n)$ can be used again as the upper bound of testing and evaluation error for further generation of optimal FST scenario set $\mathcal{X}_n$.

\subsection{Testing Scenario Set Optimization}\label{sec:fst_set_optimization}
Our remaining step is to optimize for $\mathcal{X}_n$ to achieve minimum upper bound of error. We write this optimization problem explicitly as
\begin{equation}\label{eq:objective_function_1}
\begin{aligned}
    \mathop{\min}_{\mathcal{X}_{n}}\ L_{\bm{\theta}}(\mathcal{X}_{n})& \\
    \mathrm{s.t.}\ L_{\bm{\theta}}(\mathcal{X}_{n})&=\mathop{\max}_{m\in\mathcal{M}}\left|\Tilde{\mu}_{\mathrm{FST}}-\mu_m\right|\\
    &=\mathop{\max}_{m\in\mathcal{M}}\left|\sum_{i=1}^{n}{P_m(A|\bm{x}_i)w(\bm{x}_i; \mathcal{X}_n)-\mu_m}\right|,
\end{aligned}
\end{equation}
and $w$ is given by Eq.~(\ref{eq:weight_function}). We use the classic gradient descent method to solve this optimization problem (Fig.~\ref{fig:fst_similarity_framework}). The initial FST scenario set for optimization is sampled with the same strategy $P_c$ in Section~\ref{sec:fst_similarity_learning}. 

Note again that the loss function $L_{\bm{\theta}}(\mathcal{X}_{n})$ is directly the upper bound of estimation error in Eq.~(\ref{eq:target_fst_3}). Till this section we optimize for a minimized upper bound of error through scenario similarity learning and testing scenario set optimization. The upper bound of error serves as a powerful guarantee of the accuracy of FST method. 

\subsection{Additional Error Reduction}
In this section we present an additional discussion on condition that the real AV under test $m^*\notin\mathcal{M}$, which is possible if $\mathcal{M}$ is still inaccurate. If we decompose $m^*$ into $m^* = m + \Delta m$ where $m\in\mathcal{M}$, we will have
\begin{equation}\label{eq:additional_error}
\begin{aligned}
    E^* &= \left|\Tilde{\mu}_{m,\mathrm{FST}} + \Tilde{\mu}_{\Delta m,\mathrm{FST}} - \mu_m - \mu_{\Delta m}]\right|\\
    &\leq|\Tilde{\mu}_{m,\mathrm{FST}} - \mu_m| + |\Tilde{\mu}_{\Delta m,\mathrm{FST}}| + |\mu_{\Delta m}|\\
    &\leq E_{\mathrm{max}} + |\Tilde{\mu}_{\Delta m,\mathrm{FST}}| + |\mu_{\Delta m}|,
\end{aligned}
\end{equation}
where $E^*$ is testing and evaluation error on $m^*$ and $E_{\mathrm{max}}$ is the upper bound of error given by the FST method on $\mathcal{M}$. Note that in Eq.~(\ref{eq:additional_error}) $|\mu_{\Delta m}|$ is the ground truth performance index of error model $\Delta m$ and is a constant value. $|\Tilde{\mu}_{\Delta m,\mathrm{FST}}|$ is the testing and evaluation result on $\Delta m$ with FST method and is possible to be minimized. We can expand it as
\begin{equation}\label{eq:error_model_FST}
\begin{aligned}
    \Tilde{\mu}_{\Delta m,\mathrm{FST}} &= \sum_{i=1}^{n}{\left\{P_{\Delta m}(A|\bm{x}_i)\sum_{\bm{x}\in\mathcal{X}}{S(\bm{x}_i, \bm{x}; \mathcal{X}_n)}p(\bm{x})\right\}},
\end{aligned}
\end{equation}
where $P_{\Delta m}(A|\bm{x}_i)$ is the unknown testing result on $\Delta m$ and can be minimized. As $\Delta m$ is unknown, we use a fluctuation estimator similar to \cite{li2024few} as an estimation of $\Delta m$ and $P_{\Delta m}(A|\bm{x}_i)$:
\begin{equation}\label{eq:fluctuation}
    F(\bm{x}_i; \mathcal{X}_{n})\triangleq\frac{\sum_{\bm{x}\in\mathcal{X}}{[P_m(A|\bm{x})-P_m(A|\bm{x}_i)]S(\bm{x}_i, \bm{x}; \mathcal{X}_n)p(\bm{x})}}{\sum_{\bm{x}\in\mathcal{X}}{S(\bm{x}_i, \bm{x}; \mathcal{X}_n)p(\bm{x})}}.
\end{equation}
Eq.~(\ref{eq:fluctuation}) use the difference between a FST scenario $\bm{x}_i$ and the other scenarios in scenario space weighted by the similarity and exposure frequency to estimate $P_{\Delta m}(A|\bm{x}_i)$. It is practically reasonable that if testing performance on $\bm{x}_i$ is significantly different from $\bm{x}$ with large similarity to it, there may be large potential error using the SMs on $\bm{x}_i$ and $P_{\Delta m}(A|\bm{x}_i)$ may be large.

Replacing $P_{\Delta m}(A|\bm{x}_i)$ with $F(\bm{x}_i; \mathcal{X}_{n})$ in Eq.~(\ref{eq:error_model_FST}) and we will get an approximation of $\Tilde{\mu}_{\Delta m,\mathrm{FST}}$, denoted as $\Tilde{\mu}_{F,\mathrm{FST}}$. Ignoring the constant item and assigned a weight parameter to balance the contribution of original error and additional error, the optimization target in Eq.~(\ref{eq:additional_error}) is written as
\begin{equation}\label{eq:objective_function_2}
\begin{aligned}
    \mathop{\min}_{\mathcal{X}_{n}}\ L(\mathcal{X}_{n})& \\
    \mathrm{s.t.}\ L(\mathcal{X}_{n})&=w_{\mathcal{M}}\mathop{\max}_{m\in\mathcal{M}}\left\{\left|\Tilde{\mu}_{\mathrm{FST}}-\mu_m\right|\right\}+|\Tilde{\mu}_{F,\mathrm{FST}}|.
\end{aligned}
\end{equation}
$w_{\mathcal{M}}$ is the confidence parameter we have on SM set $\mathcal{M}$. With assumption $m^*\in\mathcal{M}$, $w_{\mathcal{M}}$ is set to $\infty$. The optimization is conducted in the same way as Section~\ref{sec:fst_set_optimization}.

\section{Experiment}
\subsection{Cut-in Scenario}
\begin{figure}[ht]
  \centering
  \includegraphics[width=8.85cm]{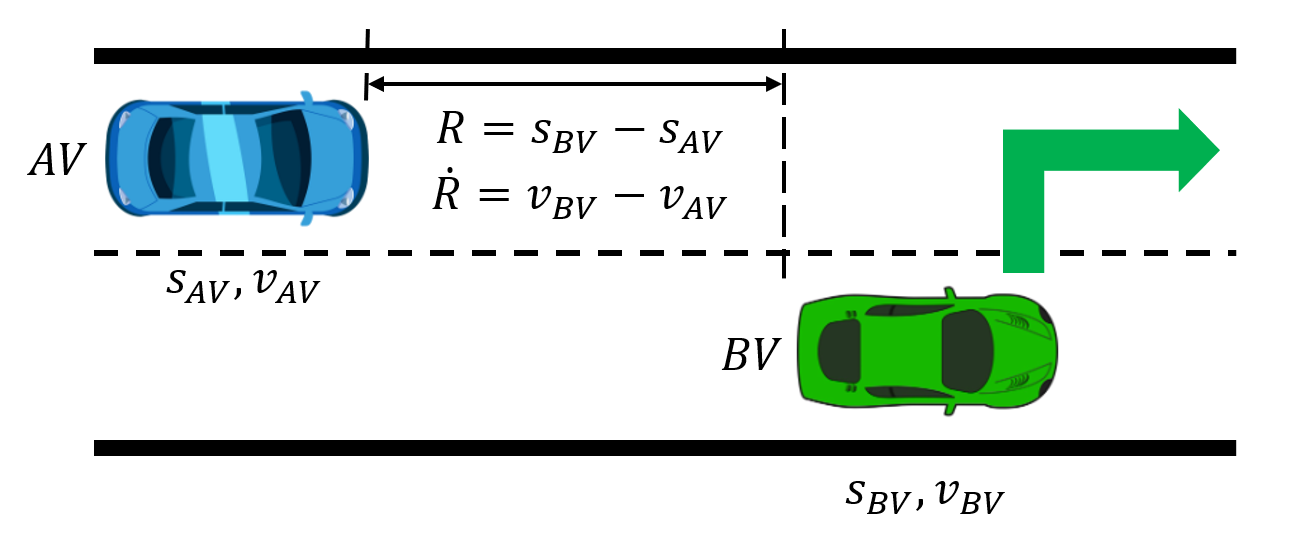}
  \caption{Cut-in scenario.}
  \label{fig:cutin}
\end{figure}
\begin{figure}[ht]
  \centering
  \includegraphics[width=8.85cm]{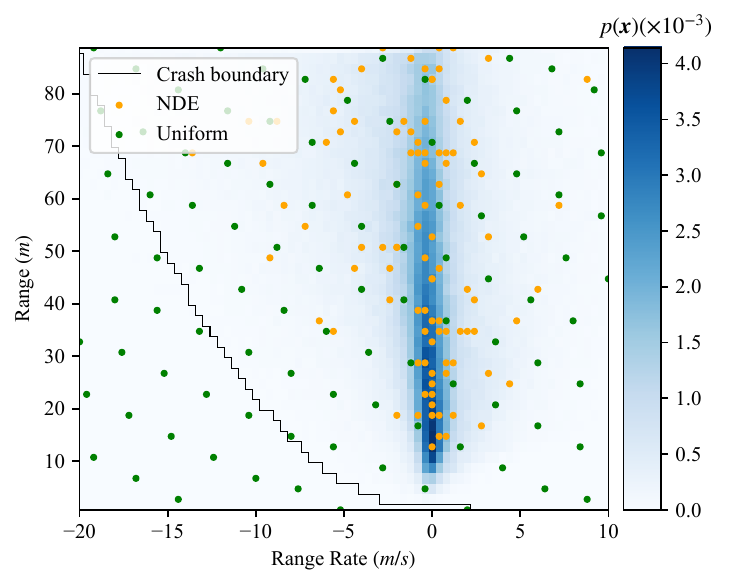}
  \caption{Illustration of cut-in state space, AV performance, and baseline testing methods.}
  \label{fig:baseline}
\end{figure}
In this section we use the simulation experiment in cut-in scenario to verify proposed FST method. Cut-in (as depicted in Fig.~\ref{fig:cutin}) is a simple and common scenario for AV testing\cite{zhao2017accelerated, zhang2018accelerated, feng2020part2}. The background vehicle (BV) changes lane before AV in this scenario, causing risk for collisions. The state space of cut-in scenario is simplified as a 2-dimensional variable
\begin{equation}
    \bm{x} = [R, \dot{R}], R\in[0,90]m, \dot{R}\in[-20, 10]m/s,
\end{equation}
where $R$ and $\dot{R}$ denote the range and range rate at the moment of lane-changing maneuver. By collecting driving behaviors of human drivers in NDE, we can derive the exposure frequency of scenarios as $p(\bm{x})$. In this experiment we focus on the crash rate and take crash event as $A$. Then if AV has collision with BV we have $P(A|\bm{x})=1$, otherwise $P(A|\bm{x})=0$. We initialize and run the cut-in scenario in simulation to get the performance of AVs under test.

The BVs in our experiment are 4 intelligent driver models (IDM) (denoted as $m_1, ..., m_4$) with different parameters. $\mathcal{M}$ is constructed according to Eq.~(\ref{eq:sm}). We use another 3 IDMs as AVs under test to verify our FST method. The crash boundary of an AV and the exposure frequency of scenarios are shown in Fig.~\ref{fig:baseline}. Scenarios with smaller $R$ and $\dot{R}$ are of higher risks and may cause crashes. It can be seen from Fig.~\ref{fig:baseline} that exposure frequency of crash scenarios in NDE is closed to 0, which results in the low efficiency or large variance of current testing methods. The crash rates of AV models in our experiment are $2.97\times10^{-3}$, $1.42\times10^{-3}$, and $6.64\times10^{-4}$, denoted as AV-1, AV-2, AV-3, respectively.
\begin{figure*}[!t]
  \centering
  \includegraphics[width=18cm]{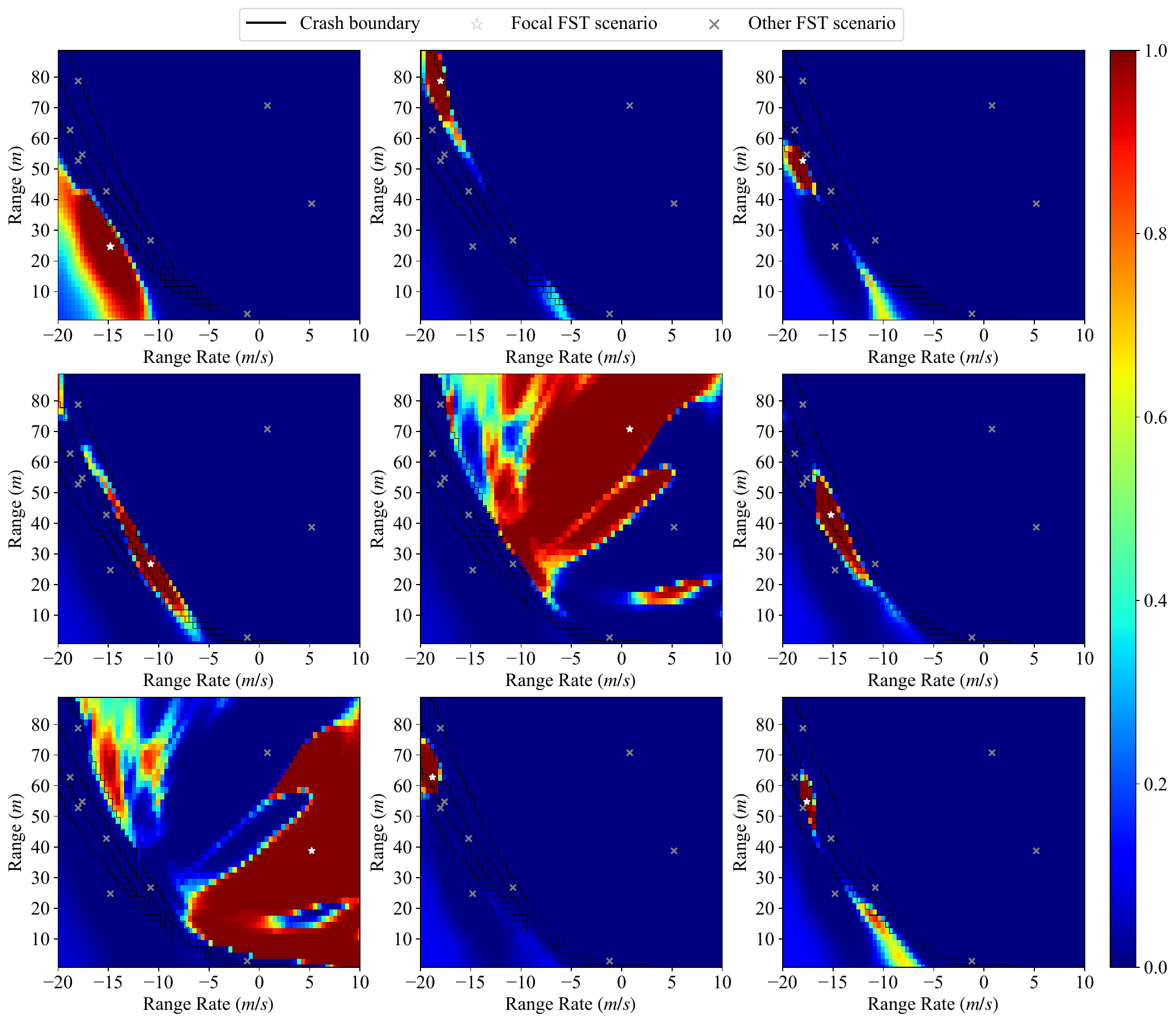}
  \caption{Examples of FST similarity map with 10 scenarios.}
  \label{fig:attn_map}
\end{figure*}

We use CMC to test AV in NDE and random quasi-Monte Carlo (RQMC)\cite{practicalqmc} for uniform testing. The results of NDE and uniform test with $100$ samples are also shown in Fig.~\ref{fig:baseline}. Scenarios in NDE gather in areas with high exposure frequency and the crash scenarios are hardly concerned. Uniform sampling is a potential way for extracting information from the state space and can lead to higher testing accuracy.

\subsection{Qualitative Analysis}

First we discuss the qualitative performance of FST method. We set the FST budget $n=10$ and train the similarity network using SM set $\mathcal{M}$. Because of the complexity of temporal simulation, the performance of AV $P_{m^*}(A|\bm{x})$ is hard to be directly represented as a linear combination of BVs and we set the optimization parameter $w_{\mathcal{M}}=1$. 

The basic idea of FST method is to extract the information of state space provided by SMs leveraging the scenario similarity. We use an example of similarity map to illustrate the similarity learned in our experiment in Fig.~\ref{fig:attn_map}. The crash boundaries of 4 SMs are shown in the black curve. We use colors to represent for the normalized similarity between a FST scenario and the scenario space, i.e. $S(\bm{x}_i, \bm{x}; \mathcal{X}_n)$. As the similarity of certain $\bm{x}_i$ changes with $\mathcal{X}_n$, we inspect different scenarios in the FST scenario set and draw its relative similarity map. For FST scenarios in the safe and high-frequency region, the weight $w(\bm{x}; \mathcal{X}_n)$ is large because of a generally large similarity value in the safe state space. In scenarios with high crash risks and where SMs perform differently, the testing scenarios are assigned less weight. Additionally, the subtle performance differences near the crash boundaries of SMs can also be reflected in the similarity of scenarios.

\begin{figure*}[h]
  \centering
  \includegraphics[width=18cm]{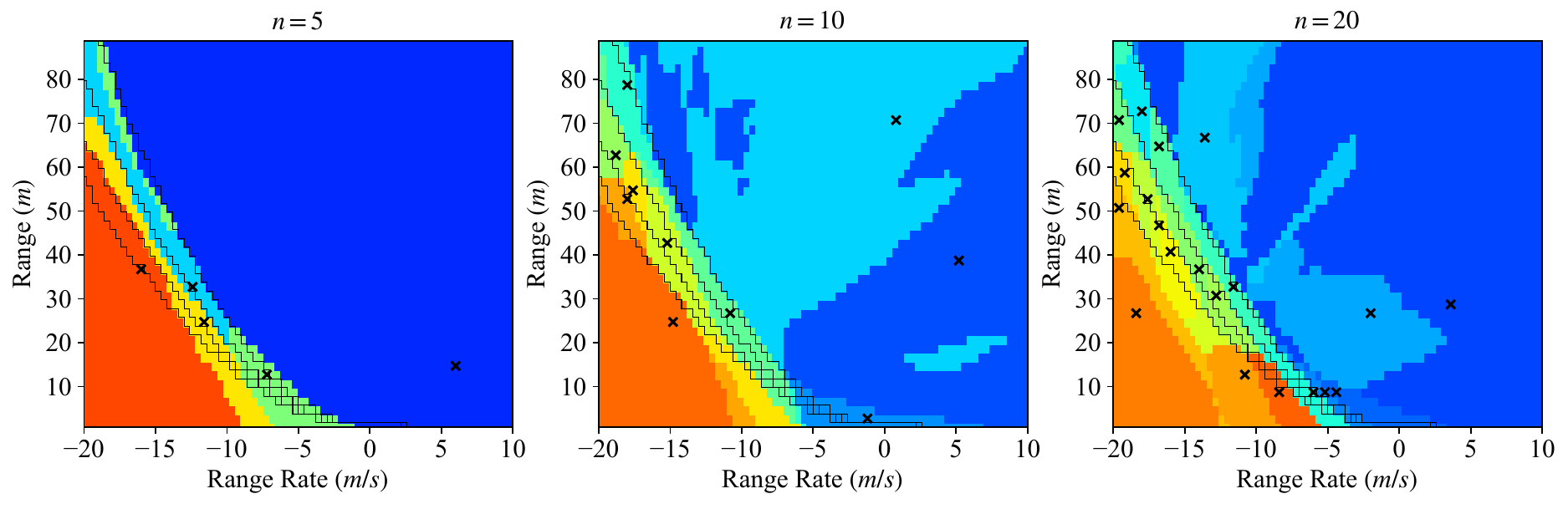}
  \caption{Example of FST scenarios and the relative similarity. The color of a scenario in state space is determined by the FST scenario with the maximum similarity.}
  \label{fig:division}
\end{figure*}

For further verification of FST method, We apply the same similarity network on testing tasks with $n=5, 10, 20$ and search for the optimal FST scenario set. Fig.~\ref{fig:division} shows the generation results of the FST scenario set and the relative similarity of scenarios. In all three cases with different testing budgets, only a small percentage of testing scenarios are in high-probability safe scenarios, which means sufficient information in this area is gained with small number of scenarios. As a contrast, most testing resources are allocated to areas near the crash boundaries. In these areas, the performance of potential AV is supposed to be more uncertain, thus in need of more tests to get accurate results. We assigned colors to scenarios in the state space according to the maximum similarities they have to FST scenarios. Remarkably, the dividing lines between FST scenarios is close to the accident boundary of SMs, indicating that the structure of SMs' state space can be learned with our method.

\subsection{Quantitative Testing and Evaluation Results}
\begin{figure*}[h]
  \centering
  \includegraphics[width=18cm]{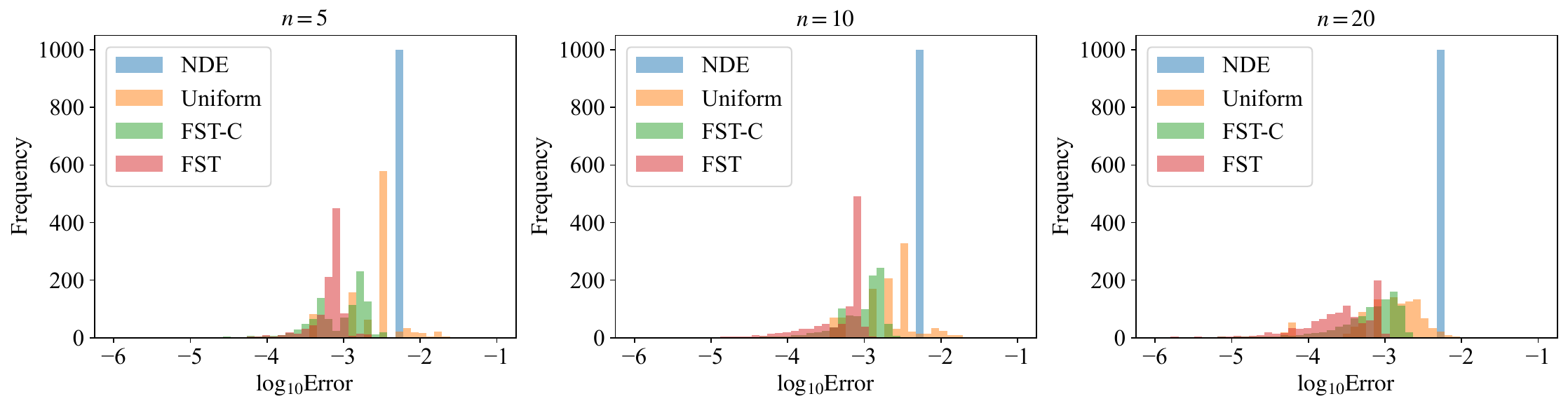}
  \caption{The comparison of distributions of estimation error on AV-1.}
  \label{fig:main_comparison_result}
\end{figure*}

In order to quantify the efficiency the FST method, we use NDE testing, uniform testing and the previous coverage-based FST (FST-C)\cite{li2024few} as baselines to carry out tests on AVs. As the testing and evaluation result is deterministic with the same FST scenarios set, we introduce randomness to our method by randomly initializing scenarios for optimization. We set $n=5, 10, 20$, which are all extremely small numbers of tests. The distributions of logarithmic evaluation error on AV-1 for these methods are shown in Eq.~(\ref{fig:main_comparison_result}). Testing are repeated 1000 times with each method. When using FST method for realistic tests, only one set of FST scenarios is adopted, so the stability of optimization can also be verified in this experiment. With NDE, most crash rates tested out of 1000 results are 0, yielding an evaluation error of $2.97\times10^{-3}$, which is totally meaningless and is the crash rate of AV-1. FST method achieves significant improvement in accuracy compared with the other methods. Furthermore, the frequency of error for FST method shows a sharp descent when the error increases, which also reflects the effectiveness of minimizing upper bound of error.

\begin{table*}[ht]
  \centering
  \caption{Statistics of testing AV-$1\sim3$ with number of samples $n = 5, 10, 20$}
  \label{tab:result_main}
  \begin{threeparttable}
  \begin{tabular}{ccccccccccc}
    \toprule
    \multicolumn{2}{c}{\multirow{2}{*}{Method}} & \multicolumn{3}{c}{Average error $(\times10^{-3})\downarrow$} & \multicolumn{3}{c}{Variance $(\times10^{-6})\downarrow$} & \multicolumn{3}{c}{Maximum error $(\alpha=1\%)$$(\times10^{-3})\downarrow$} \\
    \cmidrule(lr){3-5}\cmidrule(lr){6-8}\cmidrule(lr){9-11} && $n=5$ & $n=10$ & $n=20$ & $n=5$ & $n=10$ & $n=20$ & $n=5$ & $n=10$ & $n=20$ \\
    \midrule
    \multirow{4}{*}{AV-1} &
    NDE & $5.85/197\%$ & $5.77/194\%$ & $5.60/188\%$ & $592$ & $296$ & $148$ & $197/-$\tnote{*} & $97.0/-$ & $47.0/-$ \\
    & Uniform & $3.34/113\%$ & $2.88/96.9\%$ & $1.75/58.7\%$ & $21.3$ & $15.5$ & 4.79 & $18.7/630\%$ & $14.5/489\%$ & $6.19/208\%$\\
    & FST-C & $1.14/38.2\%$ & $1.07/36.2\%$ & $0.82/27.8\%$ & $1.77$ & $1.43$ & $0.94$ & $2.97/100\%$ & $2.15/72.4\%$ & $1.97/66.4\%$ \\
    & FST & $\bm{0.72/24.2\%}$ & $\bm{0.61/20.4\%}$ & $\bm{0.37/12.6\%}$ & $\bm{0.59}$ & $\bm{0.46}$ & $\bm{0.22}$ & $\bm{1.87/63.2\%}$ & $\bm{1.06/35.7\%}$ & $\bm{0.93/31.3\%}$ \\
    \midrule
    \multirow{4}{*}{AV-2} &
    NDE & $2.82/199\%$ & $2.80/197\%$ & $2.76/194\%$ & $284$ & $142$ & $71.0$ & $1.42/100\%$ & $98.6/-$ & $48.6/-$ \\
    & Uniform & $1.77/124\%$ & $1.42/99.9\%$ & $0.88/61.8\%$ & $6.29$ & $3.78$ & $1.49$ & $10.2/718\%$ & $6.90/485\%$ & $5.66/398\%$ \\
    & FST-C & $0.76/53.7\%$ & $0.57/40.3\%$ & $0.46/32.6\%$ & $1.06$ & $0.66$ & $0.38$ & $2.37/167\%$ & $2.34/164\%$ & $2.01/141\%$ \\
    & FST & $\bm{0.60/42.1\%}$ & $\bm{0.38/27.0\%}$ & $\bm{0.20/13.9\%}$ & $\bm{0.39}$ & $\bm{0.21}$ & $\bm{0.073}$ & $\bm{1.10/77.3\%}$ & $\bm{0.83/58.7\%}$ & $\bm{0.74/52.0\%}$ \\
    \midrule
    \multirow{4}{*}{AV-3} &
    NDE & $1.32/199\%$ & $1.32/199\%$ & $1.31/197\%$ & $132$ & $66.4$ & $33.2$ & $0.66/100\%$ & $0.66/100\%$ & $49.3/-$ \\
    & Uniform & $1.08/162\%$ & $0.77/116\%$ & $0.50/75.7\%$ & $2.47$ & $1.06$ & $0.38$ & $6.00/903\%$ & $4.33/652\%$ & $1.83/276\%$ \\
    & FST-C & $0.66/100\%$ & $0.50/74.8\%$ & $0.33/50.2\%$ & $1.39$ & $0.39$ & $0.20$ & $2.21/333\%$ & $1.77/266\%$ & $1.35/203\%$ \\
    & FST & $\bm{0.23/34.6\%}$ & $\bm{0.20/30.8\%}$ & $\bm{0.15/23.1\%}$ & $\bm{0.071}$ & $\bm{0.049}$ & $\bm{0.029}$ & $\bm{0.63/95.2\%}$ & $\bm{0.40/59.9\%}$ & $\bm{0.28/42.8\%}$ \\
    \bottomrule
  \end{tabular}
  \begin{tablenotes}
        \footnotesize
        \item[*] The relative error is too large and meaningless thus omitted.
      \end{tablenotes} 
    \end{threeparttable}
\end{table*}

We used the 4 methods to test AV-$1\sim3$ with $n=5, 10, 20$ and get the average error and variance. To verify the testing accuracy in terms of maximum error, we sorted the testing and evaluation results to get the maximum error with the confidence level of $99\%$ ($\alpha=1\%$).
The statistics with the relative error to the ground truth of AVs are listed in TABLE~\ref{tab:result_main}.

NDE testing faces the problem of ``curse of rarity" and the problem is even more serious in our experiment with small $n$. To generate valid result for these metrics, huge number of tests must be taken, so we theoretically computed the metrics for NDE testing. The metrics of the other methods are from 1000 repeated tests on AVs. From Table~\ref{tab:result_main} we can see that the testing result of NDE is useless with large errors or $100\%$ relative errors (meaning the evaluation result is 0). Proposed FST method surpasses the other methods significantly in these metrics. Notably, when the number of test turns small, the accuracy of FST method in the worst case (the maximum error) is less impaired than traditional methods. The relative error of proposed FST method is more acceptable and may have realistic availability for quick and accurate tests for AVs.

\subsection{Ideal Upper Bound of Evaluation Error}
\begin{figure}[ht]
  \centering
  \includegraphics[width=8.85cm]{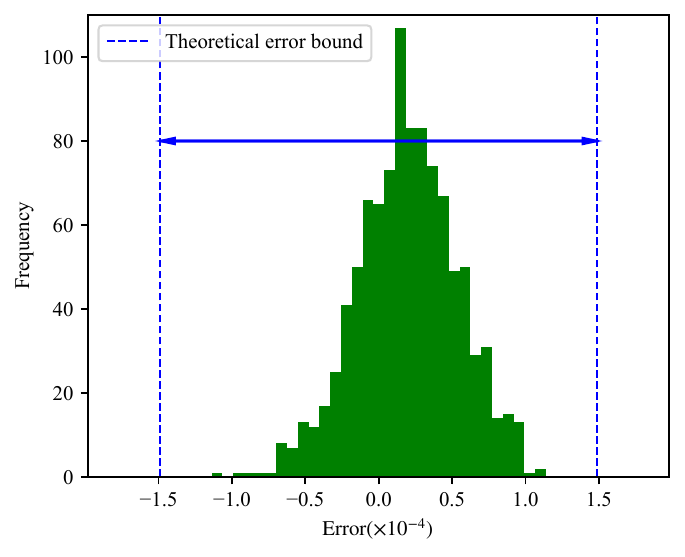}
  \caption{Illustration of upper bound of evaluation error.}
  \label{fig:error_bound}
\end{figure}

In Section~\ref{sec:method}, we proved that in ideal cases where $m^*\in \mathcal{M}$, a theoretical upper bound of evaluation error is ensured. Here we set $w_\mathcal{M}=\infty$ and search for a set of 10 optimal FST scenarios. The optimized upper bound of error is $1.49\times10^{-4}$. We manually sampled AVs from the prior knowledge set of SMs $\mathcal{M}$ for testing. The crash rate of AVs varies from $4.62\times10^{-4}$ to $4.90\times10^{-3}$. These AVs are not real enough but represent for the situations that the prior knowledge is relatively accurate. 1000 AVs are used to get the testing and evaluation error, which is shown in Fig.~\ref{fig:error_bound}. It can be seen that the evaluation error of all AVs is definitely restricted within the upper bound of error, and the maximum relative error for AV with smallest crash rate is $32.2\%$.

\subsection{Ablation Study}
\begin{table*}[ht]
  \centering
  \caption{Ablation on FST with $n=10$ samples. Without optimization module we randomly initial FST scenarios for test. $w_{\mathcal{M}}$ is set to $1$ if using fluctuation estimator module and otherwise $w_{\mathcal{M}}=\infty$. We use the coverage-based method in \cite{li2024few} (i.e. FST-C) when similarity network is not used.}
  \label{tab:ablation}
  \begin{tabular}{ccccccccc}
    \toprule
    \multirow{2}{*}{\shortstack{Similarity\\Network}} & \multirow{2}{*}{\shortstack{Optimization\\module}} & \multirow{2}{*}{\shortstack{Fluctuation\\estimator}}  & \multicolumn{3}{c}{Average error $(\times10^{-4})\downarrow$} & \multicolumn{3}{c}{Maximum error $(\alpha=1\%)$$(\times10^{-4})\downarrow$} \\
    \cmidrule(lr){4-6}\cmidrule(lr){7-9} &&& AV-1 & AV-2 & AV-3 & AV-1 & AV-2 & AV-3 \\
    \midrule
    $\times$ & $\checkmark$ & $\checkmark$ & $10.7/36.2\%$ & $5.73/40.3\%$ & $4.97/74.8\%$ & $21.5/72.4\%$ & $23.4/164\%$ & $17.7/266\%$\\
    $\checkmark$ & $\times$ & $\times$ & $14.9/50.2\%$ & $7.84/55.1\%$ & $4.37/65.8\%$ & $29.7/100\%$ & $19.0/133\%$ & $18.3/275\%$\\
    $\checkmark$ & $\checkmark$ & $\times$ & $\bm{5.64/19.0\%}$ & $3.87/27.2\%$ & $2.13/32.0\%$ & $11.5/38.7\%$ & $8.59/60.4\%$ & $5.27/79.3\%$\\
    $\checkmark$ & $\checkmark$ & $\checkmark$ & $6.06/20.4\%$ & $\bm{3.84/27.0\%}$ & $\bm{2.05/30.8\%}$ & $\bm{10.6/35.7\%}$ & $\bm{8.34/58.7\%}$ & $\bm{3.98/59.9\%}$\\
    \bottomrule
  \end{tabular}
\end{table*}

We studied the effect of different modules of FST by testing AV$1\sim 3$ with $n=10$. We gradually added the optimization module and the additional error reduction module with fluctuation estimator to verify the performance of FST method. The results are listed in Table~\ref{tab:ablation}. By combining similarity network and optimization strategy, the performance of FST is significantly improved compared with FST-C and randomly initialized FST scenarios. The effects of fluctuation estimator depends on specific AVs. Generally, the maximum evaluation error of FST method can be improved with the additional error reduction scheme.

\subsection{Similarity Network Analysis}

The similarity network is designed to learn the latent feature of scenarios in the state space to assigned weights to FST scenarios. It is possible that this feature is not strictly related to the number of tests $n$. We trained the similarity network using data consisting of $n=5, 10, 20$ scenarios respectively and conducted crossed experiments on AV-1 using $n=5, 10, 20$. The results are shown in TABLE~\ref{tab:similarity_generalization}. Training with small FST scenario set $n=5$, it is more difficult to learn accurate similarity among scenarios. With larger number of training FST scenarios, the similarity network would be applicable for different testing numbers, marking the generalization ability of FST method on different scale of testing budgets.

\begin{table}[h]
  \centering
  \caption{Generalization ability of similarity network on training and testing dataset with different $n$.}
  \label{tab:similarity_generalization}
  \begin{tabular}{cccc}
    \toprule
    \shortstack{Testing\\$n$} & \shortstack{Training\\$n$} & \shortstack{Average error\\$(\times10^{-4})\downarrow$} & \shortstack{Maximum error\\$(\alpha=1\%)$$(\times10^{-4})\downarrow$} \\
    \midrule
    \multirow{3}{*}{5} & 5 & $7.75/26.1\%$ & $19.0/64.0\%$ \\
    & 10 & $7.18/24.2\%$ & $18.8/63.2\%$ \\
    & 20 & $\bm{7.02/23.6\%}$ & $\bm{16.6/55.8\%}$ \\
    \midrule
    \multirow{3}{*}{10} & 5 & $6.67/22.5\%$ & $18.9/63.8\%$ \\
    & 10 & $\bm{6.06/20.4\%}$ & $10.6/35.7\%$ \\
    & 20 & $6.11/20.6\%$ & $\bm{9.49/32.0\%}$ \\
    \midrule
    \multirow{3}{*}{20} & 5 & $3.81/12.8\%$ & $9.97/33.6\%$ \\
    & 10 & $3.74/12.6\%$ & $9.31/31.3\%$ \\
    & 20 & $\bm{3.31/11.1\%}$ & $\bm{8.91/30.0\%}$ \\
    \bottomrule
  \end{tabular}
\end{table}

\section{Conclusion}
In this paper, we propose the few-shot testing method to tackle the challenge of testing the performance index of AVs under the limit of an extremely small testing budget. Current testing methods suffer from a low accuracy and efficiency given small number of tests, making it practically impossible to quickly obtain accurate testing and evaluation results. We deal with this problem from the perspective of searching for fixed few-shot testing scenario set to eliminate uncertainty resulted from small testing numbers and minimizing the upper bound of evaluation error. A similarity network is used to learn the feature of scenario space with surrogate models. The results show that our proposed method achieves significantly higher accuracy in few-shot testing cases and generate for the first time a practically acceptable error bound with certain confidence level. This would bring possibility for reliable quick tests for AVs in real world. Currently the efficiency of similarity calculation would increase with a larger testing state space, which is an important research direction in the future.

\bibliographystyle{IEEEtran}
\bibliography{IEEEabrv,reference.bib}


\end{document}